\documentclass[a4paper,12pt]{article}
\pdfoutput=1 

\usepackage[hmargin=.7in,vmargin=1.1in]{geometry}
\usepackage{indentfirst}
\usepackage[mathcal]{euscript}
\usepackage{amsfonts}
\usepackage{mathrsfs}
\usepackage{amsmath}
\usepackage{amssymb}
\usepackage{authblk}
\usepackage{cite}
\usepackage{xcolor}
\usepackage{mathtools}
\usepackage{tensor}
\usepackage{physics}
\usepackage{graphicx}
\usepackage{bm}
\usepackage{upgreek}
\usepackage{braket}
\usepackage{color,soul}
\usepackage{csquotes}
\usepackage{caption}
\usepackage{subcaption}
\usepackage{booktabs}
\usepackage[bookmarksnumbered=true,bookmarksopen=true]{hyperref}
\usepackage{orcidlink}
\usepackage{verbatim}

\hypersetup{
            colorlinks,
            linkcolor=[rgb]{0,0.3,0.6}, 
            citecolor=[rgb]{0,0.3,0.6}, 
            urlcolor=[rgb]{0,0.3,0.6}
           }

\def\be{\begin{eqnarray}}
\def\ee{\end{eqnarray}}

\definecolor{revred}{rgb}{0.55,0.0,0.0}

\begin{document}

\title{\Large\textbf{
 Slowly rotating condensate dark stars beyond the mean-field approximation}}

\author[a]{
Grigoris Panotopoulos  \orcidlink{0000-0002-7647-4072}
\thanks{e-mail: 
\href{mailto:grigorios.panotopoulos@ufrontera.cl}
{\nolinkurl{grigorios.panotopoulos@ufrontera.cl}}}
}

\author[b,c 
]{
\'Angel Rinc\'on {\orcidlink{0000-0001-8069-9162}} \thanks{e-mail: 
\href{mailto:angel.rincon@physics.slu.cz}
{\nolinkurl{angel.rincon@physics.slu.cz}}}
}

\author[d
]{
Ilidio Lopes  \orcidlink{0000-0002-5011-9195}
\thanks{e-mail: 
\href{mailto:ilidio.lopes@tecnico.ulisboa.pt}
{\nolinkurl{ilidio.lopes@tecnico.ulisboa.pt}}}
}

\affil[a]{\normalsize{\em Departamento de Ciencias F{\'i}sicas, Universidad de La Frontera, Casilla 54-D, 4811186 Temuco, Chile.}}

\affil[b]{\normalsize{\em
Departamento de F{\'i}sica, Universidad del B{\'i}o-B{\'i}o,
Casilla 5-C, Concepci{\'o}n, Chile.}
}

\affil[c]{\normalsize{\em Research Centre for Theoretical Physics and Astrophysics, Institute of Physics, Silesian University in Opava, Bezručovo náměstí 13, CZ-74601 Opava, Czech Republic.}}

\affil[d
]{\normalsize{\em Centro de Astrof{\'i}sica e Gravita{\c c}{\~a}o, Departamento de F{\'i}sica, Instituto Superior T{\'e}cnico-IST, Universidade de Lisboa-UL, Av. Rovisco Pais, 1049-001 Lisboa, Portugal.}}

\date{ }

\maketitle

\begin{abstract}

\smallskip\noindent

We investigate rotational properties and universal relations of slowly rotating Bose-Einstein condensate dark stars in the context of General Relativity, both at the mean-field level and when the leading beyond-mean-field Lee-Huang-Yang correction is retained self-consistently. Adopting the polytropic $n=1$ equation of state appropriate to a dilute, self-interacting Bose gas, parameterised by the boson mass $m$ and the $s$-wave scattering length $a_s$, we integrate the Tolman-Oppenheimer-Volkoff equations together with Hartle's dipole equation for the frame-dragging angular velocity, and we compute the moment of inertia, the gravito-electric tidal Love number and the dimensionless tidal deformability. The resulting equilibrium sequences yield gravitational masses in the $1$--$2\,M_{\odot}$ range with radii of $10$--$20\,\mathrm{km}$, squarely within the window presently probed by NICER and the LIGO-Virgo-KAGRA network. 
We observe that the LHY term produces a measurable reduction of the dimensionless moment of inertia at fixed compactness, whilst the I-$\Lambda$ universal relation is preserved to within a few per cent. 
We supply polynomial fits for the I-$\Lambda$ and I-$C$ relations, and show that the LHY footprint is large enough to serve as a clean diagnostic of beyond-mean-field quantum physics in a putative dark star population, complementing existing dark matter constraints from pulsar masses and from the equation-of-state interpretation of the unusually light compact remnant HESS~J1731-347.
\end{abstract}


\tableofcontents

\newpage

\section{Introduction}
\label{sec:intr}

Compact astrophysical objects have long served as natural laboratories for physics at densities unattainable in the terrestrial setting. Whilst the canonical picture treats these endpoints as neutron stars composed of strongly interacting baryonic matter, supported against gravitational collapse by degeneracy pressure and nuclear repulsion \cite{Shapiro:1983du,2000csnp.conf.....G,Haensel2007,2012ARNPS..62..485L,2025RvMP...97d5007C}, the precise composition of the core is, and remains, the great open question. A wide spectrum of exotic possibilities has been canvassed in the literature, ranging from hyperonic and kaon-condensed matter to deconfined quark phases \cite{2004Sci...304..536L,2016PhR...621..127L,2021ARNPS..71..433L}; amongst these, a particularly appealing class of models posits that the stellar fluid is, in whole or in part, a self-gravitating Bose-Einstein condensate (BEC) of dark matter particles \cite{2007JCAP...06..025B,2012PhRvD..86f4011C,2012JCAP...06..001L}, a lineage that may be traced back to the self-interacting boson stars of Colpi, Shapiro and Wasserman~\cite{Colpi:1986ye}. It is reasonable to suppose that, if the dark sector contains ultralight or weakly-interacting bosons, gravitational clustering during cosmological evolution will drive them into a condensate phase, the densest realisation of which is a gravitationally bound star.

\smallskip

Within the mean-field Hartree approximation, and in the Thomas-Fermi regime in which the quantum kinetic term is subdominant, the Gross-Pitaevskii-Poisson system for a dilute, repulsive Bose gas yields a remarkably clean result: the equation of state is polytropic with index $n=1$, namely $p = K \rho^2$ with $K = 2\pi a_s / m^3$, where $m$ is the boson mass and $a_s$ the $s$-wave scattering length \cite{2011PhRvD..84d3531C}. It transpires that, for parameter choices $(m, a_s)$ appropriate to weakly-interacting sub-GeV dark matter, this mean-field EoS supports self-gravitating configurations of a few solar masses and radii in the $10$--$20\,\text{km}$ band \cite{2012PhRvD..86f4011C}, a window further narrowed by the bosonic-dark-matter mass constraints $m \sim 0.05$--$0.5\,\text{GeV}$ derived from hybrid neutron-star models by Buras-Stubbs and Lopes \cite{2024PhRvD.109d3043B}, and it coincides, reassuringly, with the regime probed by contemporary NICER and LIGO-Virgo-KAGRA observations \cite{Miller:2021qha,Riley:2021pdl,Choudhury:2024xbk}, most notably through the tidal-deformability bound delivered by the binary neutron-star merger GW170817~\cite{LIGOScientific:2017vwq,LIGOScientific:2018hze}. Needless to say, this places BEC dark stars in a position where they may be either confirmed or ruled out by the very observational campaigns currently underway.

\smallskip

The mean-field description is not, however, the end of the story. Lee, Huang, and Yang \cite{1957PhRv..106.1135L} long ago derived the leading beyond-mean-field correction to the ground-state energy of a dilute Bose gas of hard spheres; the so-called LHY term is the analogue, for a bosonic fluid, of the perturbative Fock contribution in a Fermi liquid. Although small in the strict dilute limit, the LHY correction is by no means negligible: it is responsible, for example, for the stabilisation of ultracold quantum droplets in the laboratory \cite{2015PhRvL.115o5302P}, and its inclusion in the astrophysical setting has been shown recently to shift the mass-radius curves and tidal Love numbers of non-rotating condensate dark stars in a measurable fashion \cite{2026Physi...8...32P}. Whether the same LHY footprint persists and, indeed, whether it affects the moment of inertia, the I-Love-Q universal relations \cite{2013Sci...341..365Y,2013PhRvD..88b3009Y,2017PhR...681....1Y} and the frame-dragging profile once the star is set into slow rotation is the question to which we now turn.

\smallskip

To that end we adopt the classical Hartle slow-rotation formalism \cite{Hartle:1967he,Paschalidis:2016vmz}, wherein the metric is perturbed at first order in the angular velocity $\Omega$ and the single non-trivial equation - the $t\phi$ component of Einstein's equations - is solved for the frame-dragging function $\omega(r)$. The moment of inertia is then recovered as $I=J/\Omega$, where $J$ denotes the total angular momentum and is given by the integral of the interior frame-dragging profile first derived by Hartle \cite{Hartle:1967he}. The tidal response is quantified by the gravito-electric Love number $k$ and the dimensionless deformability $\Lambda$, extracted from the Postnikov-Hinderer Riccati equation with the appropriate surface correction \cite{Flanagan:2007ix,Hinderer:2007mb,2008PhRvD..77b1502F,Postnikov:2010yn,2010PhRvD..82b4016P}. The very same methodology has been exercised by the present authors for dark-energy stars \cite{2021PDU....3400885P,2020EPJP..135..856P}, a precedent that lends some measure of confidence to its application here.

\smallskip

A connection to the broader dark-matter program is worth remarking upon. The constraint that observed pulsars reach at least two solar masses has already been exploited to bound the particle mass of asymmetric dark matter \cite{2020PhRvD.102f3028I}; an analogous bound emerges, more cleanly, for BEC dark stars, because a single EoS - parameterised by $(m, a_s)$ - controls the entire family of equilibrium configurations. It is our view, moreover, that the BEC dark star picture dovetails naturally with hybrid-star scenarios in which baryonic and dark matter coexist within the same compact object \cite{2018PhRvD..97b4030L}, and thereby provides a complementary line of attack on the dark-matter puzzle.

\smallskip

A parallel strand of the dark-matter-in-stars program, operating in the same compact-object regime treated below, probes the equation of state of the unusually light remnant HESS~J1731-347~\cite{2022NatAs...6.1444D} through its radial oscillations, associated quark-matter models, and possible strange-quark-star interpretations~\cite{2023EPJC...83.1065R,2020PhRvD.101f3025S,2023ApJ...958...49S,DiClemente:2024ApJ}. Since that route probes the same density regime but a different sector of the equation of state, the compact-object analysis that follows is best viewed as complementary to, rather than redundant with, this line of enquiry.

\smallskip

The purpose of this paper is to present, for the first time to the best of our knowledge, a self-consistent treatment of slowly rotating BEC dark stars in General Relativity with the LHY correction retained throughout. Our original contribution is twofold: firstly, we obtain the moment of inertia, the frame-dragging angular velocity and the dimensionless tidal deformability for two representative LHY-corrected BEC models (Model A: $m=0.50\,\text{GeV}$, $a_s=0.10\,\text{fm}$; Model B: $m=0.45\,\text{GeV}$, $a_s=0.11\,\text{fm}$), in each case contrasted with the pure mean-field baseline; and secondly, we test the I-$\Lambda$ and I-$C$ universal relations beyond the mean-field approximation, producing calibrated polynomial fits that future observational campaigns may use directly. The paper is organised as follows: Section~\ref{GR} reviews the relativistic structure equations, Hartle's slow-rotation formalism, and the gravito-electric Love-number machinery. Section~\ref{sec:eos} introduces the polytropic BEC EoS and its LHY-corrected extension, presents the numerical mass-radius results, and supplies the fits for $\bar{I}(z=\ln\Lambda)$ and $\bar{I}(C)$. Section~\ref{sec:disc} draws our principal conclusions.

\smallskip

\section{Structure Equations and Tidal deformability}\label{GR}
 
In this section we shall review the equations governing the equilibrium and slow rotation of relativistic stars, together with the machinery required to extract the gravito-electric tidal Love number~$k$ and its dimensionless counterpart~$\Lambda$. The presentation is organised in two stages: the non-rotating configuration is treated in Sect.~\ref{stellareq}, whilst the first-order rotational perturbations of Hartle's formalism are developed in Sect.~\ref{stellareq:rot}.

\subsection{Non-rotating stars}\label{stellareq}

Here we briefly review the structure equations \cite{Tolman,OV} governing the interior of non-rotating relativistic stars in general relativity. For a static, spherically symmetric spacetime described in Schwarzschild-like coordinates $(t,r,\theta,\phi)$, the line element takes the form
$$\mathrm{d}s^2 = -e^{2 \nu(r)} \,\mathrm{d}t^2 + A(r) \,\mathrm{d}r^2 + r^2 \left(\mathrm{d}\theta^2 + \sin^2\theta \,\mathrm{d}\phi^2\right),$$
where the metric functions $  e^{2\nu(r)}  $ and $  A(r) \equiv e^{2\lambda(r)}  $ depend only on the radial coordinate $r$.
To facilitate the derivation, we proceed in two stages: first we obtain the equations for the static (non-rotating) configuration, and subsequently we incorporate the first-order corrections arising from slow rotation (treated in the following section).
As the first step, we will consider the Static case.
It is convenient to introduce the enclosed mass function $  m(r)  $ through the relation
\begin{align}
A(r)^{-1} \equiv 1 - \frac{2 m(r)}{r}.
\end{align}
Assuming that matter can be described as a perfect fluid, its energy-momentum tensor reads
\begin{align}
T^\mu_{\ \nu} = \mathrm{diag}(-\rho,\, p,\, p,\, p),
\end{align}
with $  \rho  $ the energy density and $  p  $ the pressure. The $  tt  $ and $  rr  $ components of Einstein's field equations then yield
\begin{eqnarray}
m'(r) &=& 4\pi r^2 \rho(r), \\
\nu'(r) &=& \frac{m(r) + 4\pi r^3 p(r)}{r^2 (1 - 2m(r)/r)},
\end{eqnarray}
where a prime denotes differentiation with respect to $  r  $.
Equivalently, the hydrostatic equilibrium condition follows from the conservation of the energy-momentum tensor:
$$p'(r) = -[\rho(r) + p(r)] \nu'(r).$$
Combining these relations produces the standard Tolman--Oppenheimer--Volkoff (TOV) equations \cite{Tolman,OV}:
\begin{eqnarray}
m'(r) &=& 4\pi r^2 \rho(r), \\
p'(r) &=& -[\rho(r) + p(r)] \frac{m(r) + 4\pi r^3 p(r)}{r^2 (1 - 2m(r)/r)}, \\
\nu'(r) &=& -\frac{p'(r)}{\rho(r) + p(r)}. \label{eq:nuTOV}
\end{eqnarray}
These must be supplemented by an equation of state relating $  p  $ and $  \rho  $, which will be specified later.
The system is integrated subject to the central boundary conditions
\begin{align}
m(0) = 0, \qquad p(0) = p_c,
\end{align}
and the surface conditions at $  r = R  $
\begin{align}
m(R) = M, \qquad p(R) = 0,
\end{align}
which determine the total mass $  M  $ and radius $  R  $ of the star.
Finally, the metric function $  \nu(r)  $ is obtained by integrating Eq.~(\ref{eq:nuTOV}) outward from the surface, where the value of $  \nu(R)  $ is fixed by the exterior Schwarzschild solution:
$$e^{2\nu(R)} = 1 - \frac{2M}{R}.$$
Thus,
$$\nu(r) = \nu(R) - \int_R^r \frac{p'(z)}{\rho(z) + p(z)}\, dz,$$
with
\begin{align}
\nu(R) = \frac{1}{2} \ln\left(1 - \frac{2M}{R}\right).
\end{align}

\subsection{Slowly rotating stars}\label{stellareq:rot}

Having reviewed the non-rotating case, let us now move to the next step, namely slow rotation, i.e.\ $J/M^2 \ll 1$, with $J$ being the angular momentum, or, equivalently, an angular velocity small compared with the mass-shedding (Keplerian) frequency, $\Omega \ll \Omega_K \sim \sqrt{M/R^3}$. At first order in the small parameter $a=J/M^2$ the diagonal part of the field equations is not modified, and therefore stellar masses and radii are still computed within the formalism of the previous subsection.
To describe slowly rotating configurations \cite{Gourgoulhon:2010ju,Paschalidis:2016vmz}, we adopt the following perturbed metric ansatz for the interior spacetime:
\begin{align}
\begin{split}
\mathrm{d}s^2 = &-e^{2 \nu(r)} \mathrm{d}t^2 + A(r) \mathrm{d}r^2 + r^2 (d \mathrm{\theta^2} + \mathrm{sin^2 \theta \: d \phi^2})
-2 \omega(r,\theta) r^2 \sin^2 \theta \mathrm{d}\phi \mathrm{d}t,
\end{split}
\end{align}
where the off-diagonal term encodes the effects of rotation to first order.
The relevant field equation is the $  t\phi  $ component of Einstein's equations,
\begin{align}
R^t_{\ \phi} = 8\pi T^t_{\ \phi},
\end{align}
retained at linear order in the rotational perturbation. The only non-vanishing component of the energy-momentum tensor at this order is the following
\begin{align}
T^t_{\ \phi} = (\rho + p) u^t u_\phi = (\rho + p) e^{-2\nu} (\Omega - \omega) r^2 \sin^2\theta.
\end{align}
Here, $\Omega$ denotes the constant angular velocity of the rigidly rotating fluid as measured by a comoving observer, while $  \omega(r,\theta)$ represents the angular velocity of an observer falling freely from infinity, evaluated to first order in $  \Omega  $. Their difference, $  \Omega - \omega  $, corresponds to the angular velocity of the fluid element relative to the locally non-rotating frame.
The four-velocity $  u^\mu  $ of the fluid satisfies the normalization $  u^\mu u_\mu = -1  $. At linear order its components read
\begin{align}
{\color{black}u^t} &{\color{black}= \left[-\left(g_{tt} + 2\Omega g_{t \phi} + \Omega^2 g_{\phi \phi}\right)\right]^{-1/2},}
\\
u^r &= 0,
\\
u^{\theta} &= 0,
\\
{\color{black}u^{\phi}} &{\color{black}= \Omega \, u^t.}
\end{align}

Next, it is convenient to introduce the auxiliary function
\begin{align}
\tilde{\omega}(r,\theta) \equiv \Omega - \omega(r,\theta).
\end{align}
Substituting into the linearized Einstein equation yields 
\begin{align} \label{eqperturb}
\begin{split}
& \frac{1}{r^4} \frac{\partial}{\partial r}
\Bigg[
  A^{-1/2} e^{-\nu} r^4 \frac{\partial \tilde{\omega}}{\partial r}
\Bigg]
+
\frac{A^{1/2} e^{-\nu}}{r^2 \sin^3 \theta}
\frac{\partial}{\partial \theta}
\Bigg[
\sin^3 \theta \frac{\partial \tilde{\omega}}{\partial \theta}
\Bigg]
=
 16 \pi (\rho + p)A^{1/2}e^{-\nu} \tilde{\omega} .
\end{split}
\end{align}
Following Hartle \cite{Hartle:1967he}, we expand $  \tilde{\omega}(r,\theta)  $ in vector spherical harmonics:
\begin{align}
\tilde{\omega}(r,\theta) = \sum_{\ell=1}^\infty \tilde{\omega}_{\ell}(r) \left( -\frac{1}{\sin\theta} \frac{dP_{\ell}}{d\theta} \right).
\end{align}
The radial functions $  \tilde{\omega}_{\ell}(r)  $ then satisfy
\begin{align}
\begin{split}
& \frac{1}{r^4} \frac{d}{dr}
\Bigg[
 A^{-1/2} e^{-\nu} r^4  \frac{d \tilde{\omega}_\ell}{d r}
\Bigg] 
-
\frac{A^{1/2} e^{-\nu}}{r^2}(\ell(\ell+1) - 2)
 \tilde{\omega}_{\ell}
= 
16 \pi (\rho + p)  A^{1/2} e^{-\nu} \tilde{\omega}_{\ell}.
\end{split}
\end{align}
In the exterior vacuum region, asymptotic flatness implies
\begin{align}
\tilde{\omega}_{\ell} \to \alpha r^{-(\ell+2)} + \beta r^{\ell-1}.
\end{align}
Matching to the asymptotic behavior $  \tilde{\omega}_{\ell} \to -2J r^{-3} + \Omega  $ as $  r\to\infty  $ shows that $  \tilde{\omega}_{\ell} = 0  $ for all $  \ell \geq 2  $. For the dipole ($  \ell=1  $) mode the equation simplifies to \cite{Staykov:2014mwa,Panotopoulos:2018joc}
\begin{align}
\begin{split}
\frac{1}{r^4} \frac{d}{dr}
\Bigg[ A^{-1/2} e^{-\nu} r^4 \frac{d \tilde{\omega}}{dr} \Bigg]
= 16 \pi (\rho + p) A^{1/2} e^{-\nu} \tilde{\omega},
\end{split}
\end{align}
subject to the boundary conditions
\begin{align} \label{eqomega}
\frac{d\tilde{\omega}}{dr}\Bigg|_{r=0} = 0, 
 \hspace{1cm}  \text{and}  \hspace{1cm}
\lim_{r \to \infty} \tilde{\omega}(r) = \Omega.
\end{align}
The moment of inertia $  I  $ of the star is defined as
\begin{align}
I \equiv \frac{J}{\Omega},
\end{align}
where $  J  $ is the total angular momentum. Using the interior solution for $  \tilde{\omega}  $ together with the boundary condition at infinity, one obtains the integral expression
\begin{align}
I = \frac{8\pi}{3} \int_0^R (\rho + p) e^{-\nu} A^{1/2} r^4 \left( \frac{\tilde{\omega}}{\Omega} \right) dr. 
\end{align}
In the non-rotating limit, $J=0=\Omega$, the moment of inertia may be estimated by the Newtonian expression $I=0.4\, M R^2$ \cite{Bejger:2002ty}, appropriate to a homogeneous sphere of radius $R$ and mass $M$. Once the $M$--$R$ profile is known, it is straightforward to compute the value of $I$.

It is convenient, throughout what follows, to work with the dimensionless moment of inertia $\bar{I}\equiv I/M^{3}$. In the next subsection we compute the source function $Q(r)$ and solve the associated Riccati equation for the logarithmic-derivative variable $y(r)\equiv r H'(r)/H(r)$, where $H(r)$ is the $\ell=2$ even-parity gravito-electric metric perturbation~\cite{Hinderer:2007mb,Postnikov:2010yn}; the surface value $y_R\equiv y(R)$ fixes the Love number $k$, whence the dimensionless deformability $\Lambda = (2k/3)\,C^{-5}$ follows with $C\equiv M/R$.

\subsection{Gravito-electric tidal Love numbers}

In what follows, let us consider a star located in an external gravitational potential $\Phi_{\rm ext}$, such as that produced by a companion in a binary system. In response to this field the star deforms, the dominant contribution being the development of a quadrupolar moment
\begin{equation}
Q_{ij} \equiv  \int d^3x \, \delta \rho(\vec{x}) \: \bigg(x_i x_j - \frac{1}{3} r^2 \delta_{ij}\bigg),
\end{equation}
a quantity directly proportional to the static external quadrupolar tidal field $E_{ij}$ via the relation 
\begin{equation}
Q_{ij} = - \lambda \: E_{ij},
\end{equation}
with
\begin{equation}
E_{ij} = \frac{\partial^2 \Phi_{\rm \text{ext}}}{\partial x^i \partial x^j},
\end{equation}
and the spatial indices run over $i,j=1,2,3$.
The dimensionless tidal Love number, defined conventionally as $k$ (a quadrupole Love number), contains the star's response and depends strongly on: i) its internal structure, ii) its mass, and finally iii) its equation of state. It is related to the dimensionful tidal deformability $\lambda$ and its dimensionless counterpart $\Lambda$ by means of
\begin{align}
\lambda &\equiv \frac{2}{3} k R^5,
\label{eq:Love1}
\\
\Lambda &\equiv \frac{2 k}{3 C^5},
\label{eq:Love2}
\end{align}
with $C=M/R$ being the factor of compactness. 

The expression for the Love number can be written in term of the set $\{C, K_o, y_R, P_5(C)\}$ as was noticed before \cite{Flanagan:2007ix,Hinderer:2007mb,Damour:2009vw,Postnikov:2010yn,Binnington:2009bb}
\begin{align}
K_{o} &= (1-2C)^2 \: [2 C (y_R-1)-y_R+2] ,
\\
y_R &\equiv y(r=R) ,
\\
P_5(C) &= 2 C \Bigl[ 4 C^4 (y_R+1) + 2 C^3 (3 y_R-2) + 
2 C^2 (13-11 y_R) + 3 C (5 y_R-8) -
3 y_R + 6 \Bigr] .
\end{align}
Thus, the Love number can be simply written as

\begin{align}
k &= \frac{8C^5}{5} \: \frac{K_{o}}{3  \:K_{o} \: \ln(1-2C) + P_5(C)} ,
\label{elove}
\end{align}
The auxiliary function $y(r)$ satisfies the Riccati differential equation \cite{Postnikov:2010yn}
\begin{align}
\begin{split}
r y'(r) + y(r)^2 + y(r) e^{\lambda (r)} \Bigl[1 + 4 \pi r^2 ( p(r) - \rho (r) ) \Bigr] 
+ r^2 Q(r) = 0 ,
\end{split}
\end{align}
with the central boundary condition $y(0)=2$. The source term $Q(r)$ (distinct from the quadrupolar moment tensor) is given by
\begin{align}
\begin{split}
  \displaystyle Q(r) = 4 \pi e^{\lambda (r)} \Bigg[ 
  5 \rho (r) 
  &+ 9 p(r) + \frac{\rho (r) + p(r)}{c^2_s(r)} 
  \Bigg] 
- 6 \frac{e^{\lambda (r)}}{r^2} - \Bigl[\nu' (r)\Bigr]^2 .
\end{split}
\end{align}
Here $c_s^2 \equiv dp/d\rho = p'(r)/\rho'(r)$ denotes the squared speed of sound.

It follows immediately that $k \propto (1-2C)^2$ and then the tidal Love numbers of black holes vanish identically at the horizon compactness $C=1/2$. 
This vanishing is a distinctive prediction of classical General Relativity: black holes, unlike other compact objects, exhibit precisely zero tidal deformability. As a consequence, a nonzero measurement of $k$ should be interpreted as direct evidence of a departure from the standard Kerr black-hole solution of GR. 
In the arena of gravitational-wave astronomy, the Einstein Telescope is expected to constrain the neutron-star equation of state with high precision \cite{Iacovelli:2023nbv}, while the Laser Interferometer Space Antenna will probe even more compact objects ($C > 1/3$) and will place bounds on the Love numbers of highly spinning central bodies at the level of $\sim 0.001$--$0.01$ \cite{Piovano:2022ojl}.

For a more detailed treatment of the gravito-electric tidal response, including alternative derivations and generalisations to higher multipoles, the reader is referred to~\cite{Flanagan:2007ix,Hinderer:2007mb,Damour:2009vw,Postnikov:2010yn,Binnington:2009bb}.

\section{Equation of state: models and results}\label{sec:eos}

The equation-of-state (EoS) is of central importance in compact star physics. It provides the thermodynamic description of matter at supranuclear densities and, through the TOV equations, fully determines the stellar mass, radius, and dynamical stability of the configuration. More precisely, the EoS is a non-trivial functional relation that connects the thermodynamic variables specifying the state of the system. Mathematically, it is commonly expressed as a power-series expansion of the pressure in terms of the energy density, the coefficients of which encode the deviations from idealized limiting cases and are, at the end of the day, derived from the elementary interactions among the constituent particles. In this manner, the EoS contains the essential microscopic dynamics, establishing a direct link between the observable macroscopic properties of the star and the fundamental forces governing its interior (see, e.g., \cite{Lattimer:2006xb,Oertel:2016bki} for more details and \cite{Gabbanelli:2018bhs,Tello-Ortiz:2020svg,Lopes:2019psm,Tello-Ortiz:2020nuc,Abellan:2023tft,Abellan:2020dze} for applications involving different EoS).

In what follows, we will assume a polytropic EoS in the mean field approximation according to which the EoS of a dilute ultracold bosonic system is given by a polytropic form \cite{Li:2012sg, Chavanis:2017loo}
\begin{equation}
p = K \rho^2, \; \; \; \; \; \;  K = \frac{2 \pi a_s}{m^3}.
\end{equation}
This result can be obtained through two independent routes summarized briefly here. The first follows from the Gross-Pitaevskii-Poisson system \cite{Chavanis:2017loo}
\begin{align}
   \mu \Psi &= \left[ -\frac{\nabla^2}{2 m} + m \Phi + g |\Psi|^2 \right] \Psi, \\ 
    \nabla^2 \Phi &=   4 \pi m  |\Psi|^2.
\end{align}
with $\Phi$ being the gravitational potential, under the Thomas-Fermi approximation, in which the quantum kinetic term is neglected, and for zero chemical potential $\mu=0$. In this limit, the radial gravitational potential satisfies the spherical Bessel equation. For spherically symmetric configurations ($  \ell=0  $), the solution is the spherical Bessel function of the first kind of order zero \cite{Abramowitz},
\begin{align}
\Phi(r) &= A \, {\color{black}j_0(r/a)} = A \frac{\sin(r/a)}{r/a}, \qquad a^2 = \frac{K}{2 \pi},
\\
\rho(r) &= m |\Psi|^2 = -\frac{m^2}{g} \Phi(r),
\end{align}
where $  A  $ is an arbitrary integration constant. Consequently, the enclosed mass function, being proportional to the gravitational potential, follows the same $  {\color{black}j_0(r/a)}  $ profile.
Equivalently, in the non-relativistic Newtonian limit, the structure equations for a self-gravitating fluid with a polytropic equation of state of index $  n=1  $ yield a density distribution that is also proportional to $  j_0(r)  $ with exactly the same scale parameter $  a  $. This constitutes one of the known exact analytic solutions of the Lane-Emden equation \cite{Chandra, Shapiro:1983du}.

\smallskip

An alternative derivation employs the Matsubara formalism and the diagrammatic expansion of many-body perturbation theory within the Hartree approximation. A pedagogical account of this route, including the resummation of ring diagrams that yields the Lee-Huang-Yang term as the first correction to the Hartree result, is given in the standard many-body treatise of Fetter and Walecka \cite{textbook}. The self-energy is found to be $  \Sigma = -g n  $, which implies $  \mu = g n  $. In the grand canonical ensemble, the Landau potential is defined as \cite{Pathria}
\begin{align}
\Omega &= E - TS - \mu N,
\end{align}
and satisfies $  \Omega = -p V  $, where $  p  $ is the pressure. At zero temperature the thermodynamic relations read \cite{Pathria}

\begin{align}
n &= \frac{\partial p}{\partial \mu}, \\
u &= \mu n - p.
\end{align}
Substituting $  \mu = g n  $ immediately yields
\begin{align}
p = u = \frac{g n^2}{2}.
\end{align}
Within the Thomas-Fermi regime the kinetic contribution is negligible, so the total energy of the system is dominated by the mean-field interaction energy.

In the astrophysical setting specifically, the non-rotating mass-radius curves reproduced in the zero-angular-velocity limit of our own integration are those of the papers~\cite{2012PhRvD..86f4011C,2026Physi...8...32P}.

We now turn to the beyond-mean-field regime. Lee, Huang and Yang~\cite{1957PhRv..106.1135L} extended the Bogoliubov theory~\cite{Bogo} by deriving the leading beyond-mean-field correction arising from quantum fluctuations; the resulting contribution is universally referred to as the Lee-Huang-Yang (LHY) term. In the dilute limit where $  n a_s^3 \ll 1  $, they obtained the following expressions for the chemical potential, energy density, and pressure:
\begin{eqnarray}
\mu & = & g n \left[ 1 + \frac{32}{3 \sqrt{\pi}} a_s^{3/2} \sqrt{n} \right], \\
\frac{E}{V} & = & \frac{1}{2} g n^2 \left[ 1 + \frac{128}{15 \sqrt{\pi}} a_s^{3/2} \sqrt{n} \right], \\
p & = & \frac{1}{2} g n^2 \left[ 1 + \frac{64}{5 \sqrt{\pi}} a_s^{3/2} \sqrt{n} \right].
\end{eqnarray}
The leading term in each expression recovers the mean-field result, while the second term represents the Lee-Huang-Yang (LHY) correction due to quantum fluctuations.

For simplicity, we introduce a dimensionless free parameter $  \zeta  $ that renders the equation of state more transparent and facilitates a direct comparison with its classical (non-quantum) counterpart. This parameter is incorporated into the EoS that includes the LHY correction. The pressure as a function of density, when quantum effects are taken into account, is then given by
\begin{align}
p &= \frac{1}{2} g n^2 \left[ 1 + \zeta
\left( \frac{64}{5 \sqrt{\pi}} a_s^{3/2} \sqrt{n} \right) \right].
\end{align}
Two limiting cases are immediately apparent: (i) $  \zeta = 0  $ recovers the purely polytropic EoS, and (ii) $  \zeta = 1  $ fully activates the LHY correction.

\begin{figure*}
	\centering
	\includegraphics[width=0.7\textwidth]{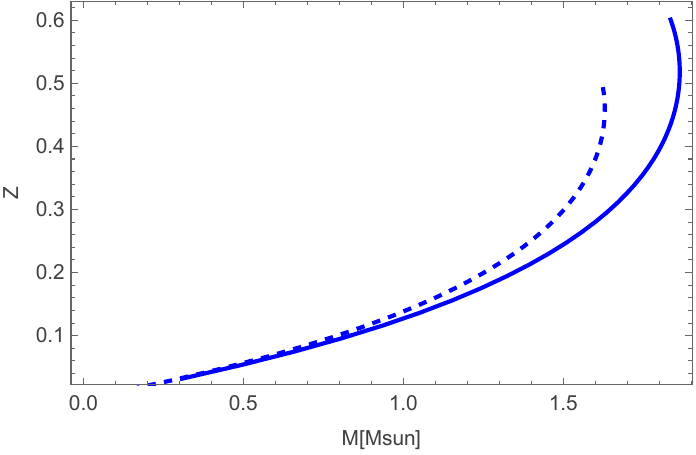} \
    \includegraphics[width=0.7\textwidth]{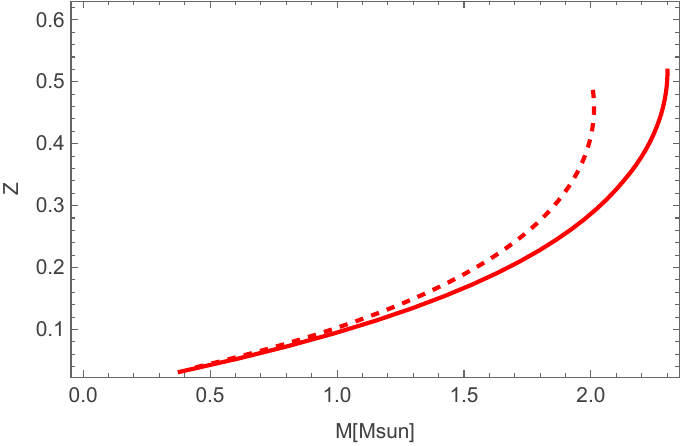}
	\caption{
	Gravitational red-shift versus stellar mass for model A (in blue color) and model B (in red color). The dashed curves correspond to the standard polytropes, while the solid curves correspond to the inclusion of the LHY term.
	}
	\label{fig:0}
\end{figure*}

In the discussion to follow we shall consider two concrete models as follows:
\begin{equation}
m=0.50 \text{ GeV}, \; \; \; \;a_s=0.10 \text{ fm}, \; \; \; \; \textrm{Model A}
\end{equation}
\begin{equation}
m=0.45 \text{ GeV}, \; \; \; \;a_s=0.11 \text{ fm}, \; \; \; \; \textrm{Model B}
\end{equation}
For these parameter choices, considered also in \cite{2026Physi...8...32P}, the equation of state supports equilibrium configurations with gravitational mass in the range $1$--$2\,M_{\odot}$ and radii in the range $10$--$20\,\mathrm{km}$, comfortably compatible with present-day NICER radius measurements~\cite{Miller:2019cac,Miller:2021qha,Choudhury:2024xbk}.
Moreover, the gravitational (or surface) red-shift, $1+Z_G = 1/\sqrt{-g_{tt}(R)}$ \cite{Hladik:2011zz}, is shown in Fig.~\ref{fig:0} as a function of the stellar mass for both models and for $\zeta=0,1$. For the metric tensor considered here the surface red-shift is computed by
\begin{equation}
Z_G = -1 + \frac{1}{\sqrt{1-2 \frac{M}{R}}}.
\end{equation}


\begin{figure*}
	\centering
	\includegraphics[width=0.8\textwidth]{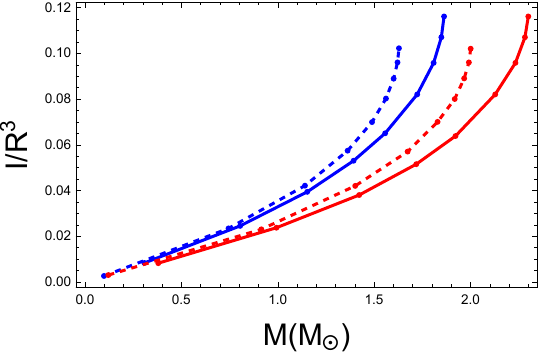}
	\caption{
		\textcolor{black}{
			Moment of inertia $I/R^3$ as a function of the stellar mass $M$ (in solar masses) for two different EoS and different values of the parameters $\{m,a_s\}$. The color code is given as follow:
              i) Dashed blue line corresponds to Model A ($\zeta=0$),
             ii) Solid blue line corresponds to Model A  ($\zeta=1$),
            iii) Dashed red line corresponds to Model B ($\zeta=0$), and finally
             iv) Solid red line corresponds to Model B  ($\zeta=1$).
    }
	}
	\label{fig:1}
\end{figure*}

Now, let us focus on the numerical fit of the moment of inertia. We make a fit first to obtain the function $Y(z)$, with $z \equiv \ln(\Lambda)$ being the independent variable, and $Y \equiv \ln(\bar{I})$ being the dependent variable, where by definition $\bar{I}=I/M^3$.
We adopt a fourth-order polynomial ansatz, as employed in the universal-relation studies of Yagi and Yunes~\cite{Yagi:2013bca,Yagi:2013awa,Yagi:2016bkt},
\begin{equation}\label{eq:YzFit}
Y(z) = a + b z + c z^2 + d z^3 + e z^4,
\end{equation}
with the coefficients $\{a,b,c,d,e\}$ determined by a least-squares fit to the numerical sequences. The resulting values, for both equations of state considered in the present work, are collected in Table~\ref{table:Second set}.

\begin{table}[ph!]
\centering
\caption{Values of the fitting coefficients of Eq.~(\ref{eq:YzFit}) for the two equations-of-state considered here, namely Models~A and B, evaluated at $\zeta=0$ (polytropic EoS) and $\zeta=1$ (beyond-mean-field approximation including the Lee-Huang-Yang correction).
}
{
\resizebox{1.0\columnwidth}{!}
{
\begin{tabular}{c|ccccc} 
\toprule
Coefficient  &  $a$  & $b$ &  $c$  &  $d$  &   $e$
\\ 
\midrule
\hline
Model A $(\zeta =0)$ & +1.54696 & $+1.36703\times 10^{-2}$ & $+3.50945\times 10^{-2}$ & $-2.18967\times 10^{-3}$ & $+7.20387\times 10^{-5}$  \\ \hline
Model B $(\zeta =0)$ & +1.38407 & $+1.43960\times 10^{-1}$ & $-1.60974\times 10^{-3}$ & $+2.09625\times 10^{-3}$ & $-1.04676\times 10^{-4}$  \\ \hline \hline
Model A $(\zeta =1)$ & +1.48497 & $+6.39624\times 10^{-2}$ & $+2.12638\times 10^{-2}$ & $-5.98423\times 10^{-4}$ & $+5.57991\times 10^{-6}$  \\ \hline
Model B $(\zeta =1)$ & +1.30982 & $+2.07751\times 10^{-1}$ & $-9.18031\times 10^{-3}$ & $+1.79536\times 10^{-3}$ & $-5.82693\times 10^{-5}$  \\ \hline
\bottomrule
\hline
\end{tabular} 
}
\label{table:Second set}
}
\end{table}

Finally, having specified the EoS to be used, we can follow \cite{Bejger:2002ty,Lattimer:2004nj} and fit the moment of inertia with the factor of compactness considering a fit of the form
\begin{equation}
I = M R^2 \: (a_2 + a_1 C + a_0 C^2),
\end{equation}
or equivalently of the form
\begin{equation}
\bar{I} \equiv \frac{I}{M^3}= a_0 + \frac{a_1}{C} + \frac{a_2}{C^2}.
\end{equation}

where the set of three parameters $\{a_0,a_1,a_2\}$ is to be determined. 
As before, the present computation considers two equations-of-state. For each EoS, two sub-cases are examined, depending on the parameters involved.

We find the following expressions for polytropic EoS:

\begin{align}
\bar{I}=
    1.06636\, 
    +
    \frac{0.188826}{C}
    +
    \frac{0.267551}{C^2},
\hspace{0.5cm}
\textrm{Model A}
\hspace{0.5cm}
\textrm{with}
\hspace{0.5cm}
\zeta =0
\\
\bar{I}=
0.689968\, 
+
\frac{0.334351}{C}
+
\frac{0.253325}{C^2}
\hspace{0.5cm}
\textrm{Model B}
\hspace{0.5cm}
\textrm{with}
\hspace{0.5cm}
\zeta =0
\end{align}

For the Lee-Huang-Yang-corrected equation of state we find, for Model A,

\begin{align}
\bar{I}=
    1.14898\,
    +
    \frac{0.208576}{C}
    +
    \frac{0.268542}{C^2}
\hspace{0.5cm}
\textrm{Model A}
\hspace{0.5cm}
\textrm{with}
\hspace{0.5cm}
\zeta =1.
\end{align}

For Model B at $\zeta=1$ we elect not to quote an $\bar{I}(C)$ polynomial fit; the corresponding LHY-corrected sequence is represented in the $Y(z)$ form of Eq.~(\ref{eq:YzFit}) with coefficients collected in the fourth row of Table~\ref{table:Second set}, which suffices for all subsequent use in the I-Love-Q analysis. The LHY correction produces, for both models, only a modest displacement relative to the mean-field baseline, consistent with the near-universality of the $I$-$\Lambda$ relation discussed above and with the I-Love-Q program of Yagi and Yunes~\cite{2013Sci...341..365Y,2017PhR...681....1Y}.

\smallskip

\begin{figure*}
	\centering
	\includegraphics[width=0.8\textwidth]{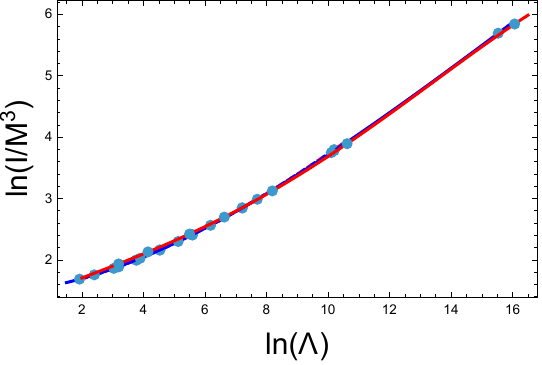}
	\caption{
		\textcolor{black}{
			 Universal relations $I-\Lambda$ for two different EoS and different values of the parameters $\{m,a_s\}$. The color code is given as follow:
              i) Dashed blue line corresponds to Model A ($\zeta=0$),
             ii) Solid blue line corresponds to Model A  ($\zeta=1$),
            iii) Dashed red line corresponds to Model B ($\zeta=0$), and finally
             iv) Solid red line corresponds to Model B  ($\zeta=1$). Bullets correspond to numerical points, whereas the continuous curves are the polynomial fits.
     }
	}
	\label{fig:2}
\end{figure*}


\begin{figure*}
	\centering
	\includegraphics[width=0.8\textwidth]{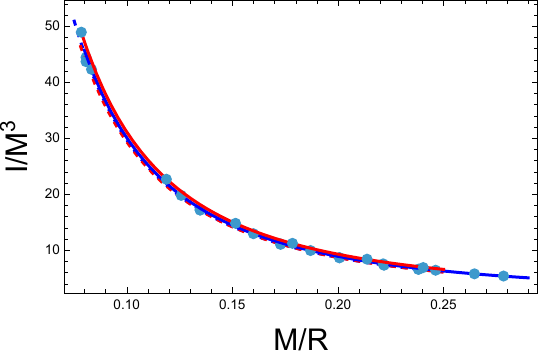}
	\caption{
		\textcolor{black}{
			 Universal relations $I-C$ for two different EoS and different values of the parameters $\{m,a_s\}$. The color code is given as follow:
              i) Dashed blue line corresponds to Model A ($\zeta=0$),
             ii) Solid blue line corresponds to Model A  ($\zeta=1$),
            iii) Dashed red line corresponds to Model B ($\zeta=0$), and finally
             iv) Solid red line corresponds to Model B  ($\zeta=1$). Bullets correspond to numerical points, whereas the continuous curves are the polynomial fits.
     }
	}
	\label{fig:3}
\end{figure*}

Figures~\ref{fig:1}, \ref{fig:2} and~\ref{fig:3} display the moment of inertia, suitably rescaled, as a function of three complementary parameters: the stellar mass $M$ (in solar units, $M_{\odot}$), the logarithm of the tidal deformability $\ln\Lambda$, and the compactness $C=M/R$. Taken together, these plots bring out the physical trends we now summarise. 

Figure~\ref{fig:1} presents the rescaled moment of inertia $I/R^{3}$ versus the stellar mass $M/M_{\odot}$ for Models~A and~B, first under the purely polytropic EoS and subsequently with the inclusion of the LHY correction to account for quantum fluctuations. For the polytropic EoS the correction introduces a pronounced modification at large masses, whilst its impact is negligible for very low-mass configurations. When the LHY correction is retained, Models~A and~B differ appreciably at high masses, in agreement with the polytropic case; at low masses, however, small but appreciable differences emerge, indicating that beyond-mean-field corrections render the moment of inertia more sensitive across the full mass spectrum.

Figure~\ref{fig:2} displays the normalised moment of inertia, $I/M^3$, versus the dimensionless deformability, $\Lambda$, in logarithmic scale for Models~A and~B under the polytropic EoS ($\zeta=0$) as well as LHY corrected one ($\zeta=1$). $\ln(I/M^{3})$ increases monotonically with $\ln(\Lambda)$. Nearly identical trends for all four cases considered in this study are observed. Bullets correspond to numerical points.

To close the picture, Figure~\ref{fig:3} presents the same normalised moment of inertia, this time versus the factor of compactness, $C=M/R$, under the same Models and the same equation-of-states. $I/M^3$ decreases monotonically with $C$, and as in the previous figure, nearly identical trends for all four cases considered in the present work are observed. Bullets correspond to numerical points. This shows that universality is not an artefact of the log-log representation, since in the $I-C$ case the plot is linear.

The physical significance of the universality of the $I-\Lambda$ relation is that it reveals an emergent behaviour of self-gravitating compact stars. Although both the moment of inertia $I$ and the tidal deformability $\Lambda$ depend sensitively on the equation of state when considered separately,
their mutual relation is remarkably insensitive to the microphysics of dense matter. This suggests that the stellar response to rotation and external tidal fields is governed mainly by the overall mass distribution and compactness rather than the detailed composition of matter in the core. When quantities are rendered dimensionless, much of the EoS dependence is absorbed into the compactness and the overall structure of the star.
Consequently, the $I-\Lambda$ relation provides a model-independent bridge between different observables. A measurement of one quantity immediately constrains the other without requiring detailed knowledge of the equation of state. This is particularly valuable in multimessenger astrophysics, where gravitational-wave measurements of $\Lambda$ can be combined with electromagnetic or pulsar observations related to $I$. The existence of such universal
relations therefore reduces the degeneracy between astrophysical observables and uncertain microphysics, allowing cleaner tests of gravity and more robust inference of compact-star properties.

\section{Discussion and final remarks}
\label{sec:disc}

In this work we have examined the equilibrium structure, slow rotation, and tidal response of self-gravitating condensate dark stars whose interior is described by the polytropic equation of state that emerges from the Gross-Pitaevskii framework once the leading beyond-mean-field, Lee-Huang-Yang correction is retained~\cite{1957PhRv..106.1135L,2015PhRvL.115o5302P,2026Physi...8...32P}. Within Hartle's slow-rotation expansion retained at first order in the angular velocity~\cite{Hartle:1967he,Paschalidis:2016vmz}, and in combination with the gravito-electric tidal formalism of Hinderer and collaborators~\cite{Hinderer:2007mb,2008PhRvD..77b1502F,2010PhRvD..82b4016P}, we have computed mass-radius sequences, moments of inertia $I$, and dimensionless tidal deformabilities $\Lambda$ as functions of the central density and of the LHY coupling strength.

Three quantitative findings stand out. First, the LHY contribution, although formally a small correction to the mean-field pressure, modifies the maximum mass and the associated radius at a level that is, whilst modest, by no means negligible; the effect is systematically controlled by the dimensionless LHY parameter introduced in Sect.~\ref{sec:eos} and reproduces, in the non-rotating limit, the results of the precursor study~\cite{2026Physi...8...32P}. Secondly, the slow-rotation solution of the dipole Hartle equation for the frame-dragging function, subject to the matching conditions summarised in Eq.~(\ref{eqomega}), delivers moments of inertia whose behaviour along each stable branch is monotonic in the gravitational mass and in quantitative agreement with the expectations set by the I-Love-Q programme~\cite{2013Sci...341..365Y,2017PhR...681....1Y}. Thirdly, the computed pairs $(I,\Lambda)$ lie on a tight, quasi-universal curve that is, within the numerical accuracy of our integration, indistinguishable from the neutron- and quark-star relations tabulated in~\cite{2017PhR...681....1Y}; this is a rather pleasing outcome, given that the microscopic physics of a dilute Bose condensate bears little superficial resemblance to that of hadronic matter.

The universal $I$-$\Lambda$ relation that we recover is, in this sense, a genuine property of the compact-object sequence rather than an artefact of the equation of state or of the log-log representation; a point that has proved useful in related programs where compact stars are admixed with, or entirely composed of, a dark component~\cite{2018PhRvD..97b4030L,2020PhRvD.102f3028I,2021PDU....3400885P,2020EPJP..135..856P}, and which has recently been extended to the rotational behaviour of exotic compact objects more generally~\cite{2026PhRvD.113d3049B}.

We therefore conclude that condensate dark stars beyond the mean-field approximation constitute a self-consistent class of compact objects whose global properties, once rotation and tidal deformation are accounted for, are compatible with current multi-messenger constraints and obey the same universal relations as their hadronic counterparts. The LHY correction, whilst quantitatively small, shifts the maximum mass and the tidal response in a predictable manner, and the approximate universality of the $I$-$\Lambda$ curve offers a practical means of constraining the underlying bosonic coupling from future measurements of moment of inertia and tidal deformability in binary coalescences.

It is worth remarking, by way of closing, that the bound on the $(m,a_s)$ plane delivered by the present compact object analysis is usefully read alongside the equation-of-state tests associated with the interpretation of the unusually light remnant HESS~J1731-347~\cite{2022NatAs...6.1444D,2023EPJC...83.1065R}. The two probes, being sensitive to rather different combinations of the microscopic parameters, are mutually complementary; their joint application narrows the admissible bosonic dark-matter parameter space more than either could achieve in isolation. This confluence, rather than any single observable, constitutes, in our view, the strongest argument for treating BEC dark stars on an equal footing with their hadronic counterparts.

\section*{Acknowledgments}

\smallskip\noindent

I.~L. thanks the Fundação para a Ciência e Tecnologia (FCT), Portugal,  for the financial support to the Center for Astrophysics and Gravitation (CENTRA/IST/ULisboa) through grant No. UID/PRR/00099/2025 (https://doi.org/10.54499/UID/PRR/00099/2025) and grant No. UID/00099/2025 (https://doi.org/10.54499/UID/00099/2025).
A.~R. would like to express his gratitude to Silesian University in Opava, Czech Republic, for their financial support.
The creation of this article was supported by the grant program Vouchers for Universities in the Moravian-Silesian Region (registration number CZ.10.03.01/00/23\_042/0000390).
This article is based upon work from COST Action FuSe, CA24101, supported by COST (European Cooperation in Science and Technology).
%

\section*{Funding Statement}
\noindent
The following grants have supported the paper: 
i) \verb|UID/PRR/00099/2025|, 
ii) \verb|UID/00099/2025|, 
iii) \verb|CZ.10.03.01/00/23_042/0000390|, and
iv) \verb|COST Action FuSe, CA24101|.

\section*{Data Availability Statement}
\noindent
The data that support the findings of this study are available from the corresponding author upon reasonable request.

\section*{Conflict of Interest Statement}
\noindent
The authors declare that they do not have any conflict of interest.

\bibliographystyle{utphys}
\bibliography{Library2_IL}

\end{document}